\documentclass[rnote,traditabstract]{aa} 
 
\usepackage{graphicx}
\usepackage{txfonts}
\usepackage{natbib}
\bibpunct{(}{)}{;}{a}{}{,}

\begin{document}

   \title{The delayed time distribution of massive double compact star mergers
         }

   \author{N. Mennekens
           \and D. Vanbeveren
          }

   \institute{Astronomy and Astrophysics Research Group, Vrije Universiteit Brussel, Pleinlaan 2, 1050 Brussels, Belgium\\
              \email{nmenneke@vub.ac.be, dvbevere@vub.ac.be}
             }

   \date{Received 26 January 2016 / Accepted March 2016}

  \abstract
   {In order to investigate the temporal evolution of binary populations in general, double compact star binaries and mergers in particular within a galactic evolution context, a most straightforward method is obviously the implementation of a detailed binary evolutionary model in a galactic chemical evolution code. To our knowledge, the Brussels galactic chemical evolution code is the only one that fully consistently accounts for the important effects of interacting binaries on the predictions of chemical evolution. With a galactic code that does not explicitly include binaries, the temporal evolution of the population of double compact star binaries and mergers can be estimated with reasonable accuracy if the delayed time distribution (DTD) for these mergers is available. The DTD for supernovae type Ia has been studied extensively the last decade. In the present paper we present the DTD for merging double neutron star binaries and mixed systems consisting of a neutron star and a black hole. The latter mergers are very promising sites for the production of r-process elements and the DTDs can be used to study the galactic evolution of these elements with a code that does not explicitly account for binaries.
}

   \keywords{binaries: close --
             stars: massive --
             Galaxy: evolution
            }
            
   \titlerunning{DTD of massive double compact star mergers}
   
   \authorrunning{Mennekens \& Vanbeveren}

   \maketitle

\section{Introduction}

The r-process nucleosynthesis is known since the pioneering works of Burbidge et al. (1957) and Seeger et al. (1965) but the main production site of the r-process elements still remains a matter of debate. Basically two sites have been studied in some detail in the recent past: type-II supernova (SN) (including long GRBs) and merging double compact star binaries, either double neutron star systems (NS+NS) or neutron star + black hole binaries (NS+BH). Type-II SN may have been main r-process sites in the early-time Galaxy when [Fe/H] $<$-3 but it is questionable that they remained important when the metallicty became large (e.g., Qian, 2011; Wanaja et al., 2011; Janka, 2012). Recent hydrodynamical simulations of the merging process of double compact star binaries have shown that during and after the merger phase some $10^{-3}$ M$_{\odot}$ up to more than 0.1 M$_{\odot}$ of neutron rich matter can be ejected from the system (e.g., Rosswog, 2005; Oechslin et al., 2007; Goriely et al., 2011; Korobkin et al., 2012; Bauswein et al., 2013; Wanajo et al., 2014; Foucart et al., 2014; Just et al., 2015). Although the results of the abovementioned simulations may be quantitatively different they all agree that the neutron rich ejecta provide the necessary conditions for very efficient r-processing.

By implementing the details of one particular r-process site in a chemical evolutionary code it is possible to compute the galactic temporal variation of r-process elements predicted by that site. Comparison with observations (mainly the observations of Eu) may yield important information on the importance of the chosen site for the overall galactic r-process production. In principle, if one decides to study in this way the effects of merging double compact star binaries, for the sake of consequence one has to combine a full binary population model (including binary chemical yields) with a galactic formation and evolution code. We like to remind the interested reader that present observations of massive stars yield a massive interacting binary frequency of at least 70\% (Sana et al., 2012) and we note that most of the existing chemical evolution codes do not account for the evolutionary consequences of binaries. Our code that was described in De Donder and Vanbeveren (2004) with important updates listed in Vanbeveren et al. (2012) and in Mennekens and Vanbeveren (2014) is to our knowledge the only code with an internal and fully self-consistent detailed binary population model (including binary yields). We used this code in order to predict the temporal evolution of the r-process elements ejected by double compact star binaries. An important conclusion resulting from our 2004 and 2014 results was that (with exception for the first say 100 Myr) compact star binary mergers may be major contributors to the r-process enrichment and that it is very plausible that this enrichment is due mainly to BH+NS mergers.

Matteucci et al. (2014) tried to evaluate the importance of double neutron star mergers as r-process production site however since they used a galactic code that does not explicitly account for binaries, approximations had to be made that may affect overall conclusions. To illustrate, the authors adopt a model with three fixed merger delay time-scales for all double compact star mergers and they correctly note that a more realistic approach would consider a distribution function of such time-scales. Moreover, since their code does not account for close binaries they have to rely on `observed' merger rates like those proposed by Kalogera et al (2004). However, these rates may tell us something about the present situation but they do not give information on the temporal (Z-dependent) evolution of the merger rates. Moreover, the rates are based on the observation of only 9-10 observed double NS binaries and are therefore very uncertain. Last not least, no observations exist on BH+NS binaries and as already discussed above merging BH+NS may be the dominant r-process production site.

Scientists that use chemo-dynamical models in order to study galactic evolution rarely have binaries included and they are requesting party for time delay distributions of merging double compact star binaries.

We find ourselves in a priviledged position in order to provide such distributions and this is the main scope of the present paper. It should not be considered as a stand alone paper but rather as an addendum to the paper Mennekens and Vanbeveren (2014, paper I).

\section{The delayed time distribution (DTD) of double compact star mergers}

The delayed time is defined as the evolutionary time ($\Delta$t$_{\mathrm{evol}}$) between the formation of a massive binary and the eventual formation of the double compact star system (NS+NS or NS+BH) + the coalescence timescale ($\Delta$t$_{\mathrm{mer}}$) of the latter due to relativistic gravitational wave emission. In our simulations $\Delta$t$_{\mathrm{evol}}$ is computed from first binary evolution principles. Note that in binaries where RLOF and mass transfer is active this evolutionary time is not simply the evolutionary time of the low mass companion of the binary. The coalescence time obviously depends on the period of the double compact star binary but it is also very much dependent on its eccentricity. To calculate the DTD it is therefore indispensable in order to calculate in detail the effects of an SN explosion (which may be asymmetric) on the binary orbital parameters. Our binary code has been introduced in De Donder and Vanbeveren (2004) and important updates were described in Vanbeveren et al. (2012) and Mennekens and Vanbeveren (2014). We invite the interested reader to consider at least the latter since the present paper is written as an addendum. Summarizing, our binary population code accounts for the following binary processes: stable RLOF in case A/Br\footnote{We follow the case A, case Br, case Bc and case C binary classification of Kippenhahn \& Weigert (1967) and Lauterborn (1970).} binaries (accounting for conservative and non-conservative assumptions characterised by the mass transfer efficiency $\beta$), common-envelope evolution in case Bc/C binaries (characterised by the energy conversion efficiency $\alpha$), common-envelope evolution of a binary where one of the components is a compact star, a detailed treatment of the effects of the supernova explosion on binary parameters (the SN can be asymmetric and the degree of asymmetry is expressed as the kick velocity that the compact star gets, also dependent on the adopted fallback mechanism as described in Fryer et al. 2012), stars that end their life in a prompt electron capture supernova (ECSN) where the resulting neutron star is formed with a small kick (Podsiadlowski et al., 2004), the formation of a BH with or without a SN explosion. Another important assumption is whether or not a luminous blue variable (LBV) phase can strip the entire envelope of a very massive ($>$ 40 M$_{\odot}$) star in a spherically symmetric way. It is obvious that the code also accounts for the details of stellar evolution which depends on e.g. stellar wind mass loss, convective core overshooting, mass accretion during stable RLOF in case A/Br binaries, the binary merger process etc.

Once a double compact star binary is formed (i.e., after the second SN explosion of the binary) our code computes the coalescence timescale using the gravitational wave radiation formalism of Landau and Lifshitz (1951). The formalism is summarized in De Donder \& Vanbeveren (2004). We repeat that the timescale very much depends on the eccentricity of the binary, and therefore a detailed model that computes the effects of an asymmetric SN explosion on binary parameters is absolutely needed.

A DTD depends on various distribution functions of star and binary parameters. In correspondence with current observations of massive stars and binaries in our Galaxy, we adopt (De Donder \& Vanbeveren 2004) a Scalo (1986) initial mass function $\sim M^{-2.7}$ for single stars and for primaries of close binaries with mass $>$ 2 M$_{\odot}$ and a Salpeter (1955) slope $\sim M^{-2.35}$ for 0.08 $\leq$ M/M$_{\odot}$ $\leq$ 2, a flat binary mass ratio distribution and a binary period distribution that is flat in the Log. Recent observations of Sana et al. (2012) support the flat mass ratio distribution however the authors conclude that a period distribution $\sim$ (Log P)$^{-0.55}$ cannot be excluded. In paper I we investigated the effect on the population of double compact star systems and concluded that it is marginal. To support the latter conclusion we will also compute the DTD with the `alternative' period distribution.

The effects of an asymmetric SN explosion on binary parameters are computed by relating the asymmetry and the kick the compact star gets. We therefore need a distribution function of possible kicks. The kick velocity distribution is linked to the observed pulsar velocity distribution and we use the study of Lyne \& Lorimer, 1994. This yields a $\chi^2$-distribution with average 450 km/s but to account for possible observational errors we also calculated the DTD using a distribution with average 265 km/s (Hobbs et al. 2005).

\section{Results}

The binary population models for which a DTD of double compact star mergers have been calculated are those described in paper I and listed in Table 1 of that paper. Since the latter concerned galactic evolution, paper I also considered various star formation rate models (which is for example the only difference between models 2 and 4), which is obviously not important here. Furthermore, only the models are withdrawn which predict a present day Galactic double neutron star merger rate\footnote{This is the same rate as calculated by e.g. Dominik et al. 2012.} $\geq$ 3/Myr which is a most probable `observational' lower limit (Kim et al., 2010). According to the results of Table 2 in paper I, the models 1, 3, 8, 9, 10, 11, 17, 18, 22 do not meet this lower limit. Model 22 predicts a merger rate of 2.5/Myr which means that rejecting this model is at first glance not based on a strong argument. An additional argument to reject model 22 comes from the predicted DTD of type Ia supernovae if the same binary parameter values are used as in model 22. The SN Ia DTD predicted by the Brussels code has been described in detail in Mennekens et al. (2010) and is shown in Figure 1 if the model 22 parameters are used. As can be noticed the theoretically predicted DTD deviates from observations by a large factor and this strengthens the conclusion to reject model 22.

\begin{figure}[]
\centering
   \includegraphics[width=8.4cm]{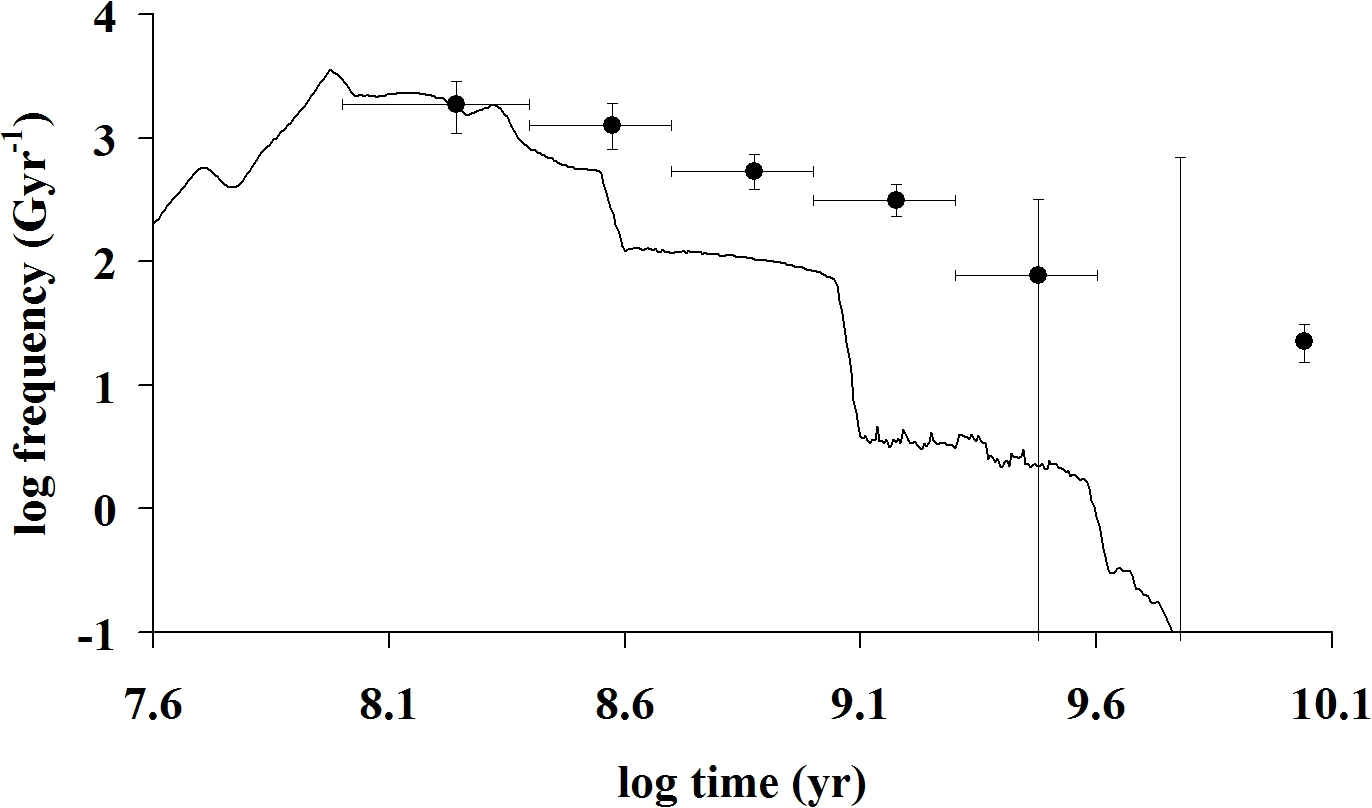}
     \caption{SN Ia DTD obtained using model 22 from Paper I (solid line), compared to observations (Totani et al. 2008; dots). See Mennekens et al. (2010) for details on computation and comparison method.}
     \label{fig:1}
\end{figure}

\begin{figure}[]
\centering
   \includegraphics[width=8.4cm]{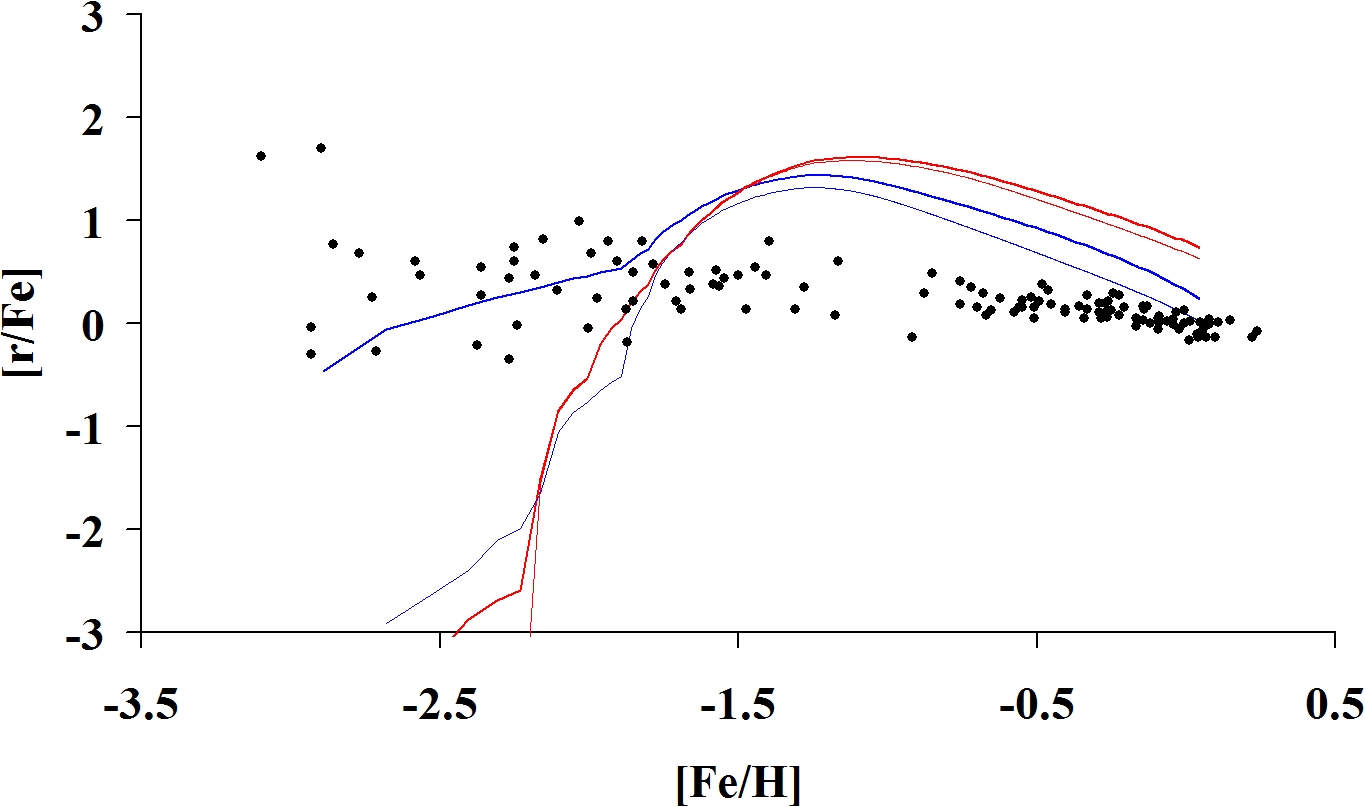}
     \caption{r-process element enrichment obtained using model 6 (blue) and 19 (red) from Paper I. The thin lines indicate enrichment taking into account only NSNS mergers. The black dots represent observations (see Mennekens \& Vanbeveren 2014).}
     \label{fig:2}
\end{figure}

Using our Galactic Chemical Evolution code that includes a detailed treatment of binaries, we also computed the predicted temporal evolution of the r-process elements for all the binary models. It is clear that the models that predict too much r-process element enrichment have to be rejected as well\footnote{Any r-process element production process that has not yet been recognized and/or has not yet been included in our simulations will obviously strengthen this conclusion.}. This is the case for models 6, 14, 15, 16, 19, 20, 21, 23. To illustrate this criterion Figure 2 shows the predicted temporal evolution of r-process enrichment for two representative binary models and we compare with observations\footnote{We use the same observations as those discussed in Paper I.}.  Interestingly, the models that do not predict too much r-process enrichment have a present combined galactic NS+NS and NS+BH merger rate $\leq$ 20/Myr ($\leq$ 10/Myr for both).

\begin{table*}
\centering
\caption{Double compact star merger DTD for Z=0.002. Values are logarithmic, in number of events per time bin after a starburst with total mass 1 M$_{\odot}$.}
\begin{tabular}{c | c c c c c c c c c c}
\hline
log $t$	& mod. 2/4 & mod. 2/4 & mod. 5 & mod. 5 & mod. 7 & mod. 7 & mod. 12 & mod. 12 & mod. 13 & mod. 13 \\
(Gyr)	&	NSNS	&	BHNS	&	NSNS	&	BHNS	&	NSNS	&	BHNS	&	NSNS	&	BHNS	&	NSNS	&	BHNS	\\
\hline
-2.05	&	 	&	-7.973	&	 	&	-7.848	&		&	-7.360	&	 	&	 	&	 	&	-8.116	\\
-1.95	&	 	&	-7.051	&	 	&	-7.005	&		&	-7.370	&	 	&	 	&	 	&	-7.190	\\
-1.85	&	-9.521	&	-6.699	&	-8.981	&	-6.632	&	-9.449	&	-7.163	&	-9.521	&	-9.229	&	-9.816	&	-6.836	\\
-1.75	&	-9.735	&	-6.515	&	-9.932	&	-6.421	&	-9.524	&	-7.230	&	-9.735	&	-10.340	&	-10.035	&	-6.650	\\
-1.65	&	-9.897	&	-6.639	&	-10.398	&	-6.610	&	-9.277	&	-6.872	&	-9.897	&	-9.849	&	-10.165	&	-6.771	\\
-1.55	&	-10.065	&	-6.976	&	-9.085	&	-6.958	&	-8.315	&	-6.855	&	-10.065	&	-8.902	&	-10.349	&	-7.121	\\
-1.45	&	-10.438	&	-7.288	&	-9.565	&	-7.271	&	-7.738	&	-6.786	&	-10.438	&	-9.132	&	-10.714	&	-7.451	\\
-1.35	&	-7.943	&	-7.460	&	-7.899	&	-7.427	&	-7.418	&	-6.911	&	-7.943	&	-9.211	&	-8.012	&	-7.636	\\
-1.25	&	-9.040	&	-7.564	&	-9.049	&	-7.528	&	-7.450	&	-6.745	&	-9.040	&	-9.331	&	-9.294	&	-7.743	\\
-1.15	&	-8.072	&	-7.608	&	-8.067	&	-7.560	&	-6.455	&	-6.681	&	-8.072	&	-9.457	&	-8.185	&	-7.783	\\
-1.05	&	-7.451	&	-7.566	&	-7.422	&	-7.502	&	-6.300	&	-6.724	&	-7.451	&	-9.915	&	-7.561	&	-7.732	\\
-0.95	&	-7.408	&	-7.441	&	-7.389	&	-7.367	&	-6.515	&	-6.731	&	-7.408	&	-9.793	&	-7.525	&	-7.595	\\
-0.85	&	-7.328	&	-7.287	&	-7.311	&	-7.213	&	-6.817	&	-6.791	&	-7.328	&	-9.702	&	-7.452	&	-7.436	\\
-0.75	&	-7.536	&	-7.209	&	-7.425	&	-7.139	&	-6.981	&	-6.794	&	-7.536	&	-9.933	&	-7.667	&	-7.351	\\
-0.65	&	-7.692	&	-7.187	&	-7.677	&	-7.127	&	-7.161	&	-6.816	&	-7.692	&	-10.494	&	-7.824	&	-7.324	\\
-0.55	&	-7.961	&	-7.198	&	-8.045	&	-7.141	&	-7.285	&	-6.812	&	-7.961	&	-10.577	&	-8.073	&	-7.327	\\
-0.45	&	-8.250	&	-7.166	&	-8.177	&	-7.168	&	-7.504	&	-6.787	&	-8.250	&	-11.083	&	-8.350	&	-7.287	\\
-0.35	&	-8.325	&	-7.251	&	-9.003	&	-7.196	&	-7.730	&	-6.763	&	-8.325	&	-9.968	&	-8.452	&	-7.363	\\
-0.25	&	-8.758	&	-7.187	&	-10.298	&	-7.217	&	-7.932	&	-6.688	&	-8.758	&	-7.546	&	-8.888	&	-7.280	\\
-0.15	&	-9.181	&	-7.250	&	 	&	-7.231	&	-8.236	&	-6.676	&	-9.181	&	-9.010	&	-9.319	&	-7.349	\\
-0.05	&	 	&	-7.251	&	 	&	-7.239	&	-8.540	&	-6.601	&	 	&	-8.005	&	 	&	-7.332	\\
0.05	&	 	&	-7.103	&	 	&	-7.234	&	-8.767	&	-6.588	&	 	&	-8.405	&	 	&	-7.171	\\
0.15	&	 	&	-6.971	&	 	&	-7.227	&	-9.308	&	-6.489	&	 	&	-8.023	&	 	&	-7.013	\\
0.25	&	 	&	-7.155	&	 	&	-7.213	&	-9.426	&	-6.436	&	 	&	-7.461	&	 	&	-7.230	\\
0.35	&	-12.049	&	-6.988	&	 	&	-7.206	&	-9.751	&	-6.562	&	-12.049	&	-7.537	&	-12.333	&	-7.043	\\
0.45	&	-10.227	&	-7.140	&	-10.249	&	-7.205	&	-9.281	&	-6.477	&	-10.227	&	-7.535	&	-10.496	&	-7.198	\\
0.55	&	-9.915	&	-7.255	&	-9.579	&	-7.202	&	-8.896	&	-6.432	&	-9.915	&	-7.294	&	-10.181	&	-7.318	\\
0.65	&	-9.966	&	-7.474	&	-9.484	&	-7.190	&	-8.414	&	-6.582	&	-9.966	&	-7.524	&	-10.214	&	-7.529	\\
0.75	&	-9.988	&	-7.529	&	-9.672	&	-7.202	&	-7.811	&	-6.503	&	-9.988	&	-7.438	&	-10.212	&	-7.577	\\
0.85	&	-10.274	&	-7.559	&	-9.988	&	-7.226	&	-7.637	&	-6.616	&	-10.274	&	-7.584	&	-10.542	&	-7.604	\\
0.95	&	-10.074	&	-7.593	&	-9.918	&	-7.266	&	-7.601	&	-6.573	&	-10.074	&	-7.547	&	-10.308	&	-7.636	\\
1.05	&	 	&	-7.665	&	-10.297	&	-7.326	&	-7.592	&	-6.707	&	 	&	-7.520	&	 	&	-7.710	\\
1.15	&	 	&	-7.758	&	-10.896	&	-7.378	&	-7.555	&	-6.717	&	 	&	-7.408	&	 	&	-7.804	\\
1.25	&	 	&	-7.895	&	-11.535	&	-7.453	&	-7.628	&	-6.666	&	 	&	-7.475	&	 	&	-7.940	\\
\hline
\end{tabular}
\label{tab:DTD_002}
\end{table*}

\begin{table*}
\centering
\caption{Double compact star merger DTD for Z=0.02. Values are logarithmic, in number of events per time bin after a starburst with total mass 1 M$_{\odot}$.}
\begin{tabular}{c | c c c c c c c c c c}
\hline
log $t$	& mod. 2/4 & mod. 2/4 & mod. 5 & mod. 5 & mod. 7 & mod. 7 & mod. 12 & mod. 12 & mod. 13 & mod. 13 \\
(Gyr)	&	NSNS	&	BHNS	&	NSNS	&	BHNS	&	NSNS	&	BHNS	&	NSNS	&	BHNS	&	NSNS	&	BHNS	\\
\hline
-1.85	&	 	&	-9.715	&	 	&	-9.715	&		&	-7.751	&	 	&	 	&	 	&	-9.857	\\
-1.75	&	 	&	-8.789	&	 	&	-8.790	&		&	-7.794	&	 	&	 	&	 	&	-8.928	\\
-1.65	&	 	&	-8.618	&	 	&	-8.641	&		&	-7.818	&	 	&	-9.813	&	 	&	-8.757	\\
-1.55	&	-9.285	&	-8.444	&	-9.282	&	-8.461	&	-7.801	&	-8.110	&	-9.285	&	-9.655	&	-9.593	&	-8.570	\\
-1.45	&	-9.667	&	-8.723	&	-9.699	&	-8.757	&	-7.881	&	-8.278	&	-9.667	&	-9.588	&	-9.975	&	-8.884	\\
-1.35	&	-9.272	&	-9.069	&	-9.272	&	-9.127	&	-8.034	&	-8.393	&	-9.272	&	-10.296	&	-9.563	&	-9.231	\\
-1.25	&	-8.754	&	-9.164	&	-8.752	&	-9.267	&	-7.967	&	-8.632	&	-8.754	&	-11.010	&	-9.027	&	-9.345	\\
-1.15	&	-6.681	&	-9.235	&	-6.681	&	-9.332	&	-6.603	&	-8.677	&	-6.681	&	-10.508	&	-6.810	&	-9.431	\\
-1.05	&	-6.170	&	-9.392	&	-6.170	&	-9.419	&	-6.320	&	-8.664	&	-6.170	&	-10.378	&	-6.297	&	-9.599	\\
-0.95	&	-6.204	&	-9.422	&	-6.204	&	-9.585	&	-6.577	&	-8.725	&	-6.204	&	-10.568	&	-6.346	&	-9.633	\\
-0.85	&	-6.368	&	-9.459	&	-6.368	&	-9.720	&	-6.848	&	-8.847	&	-6.368	&	-10.876	&	-6.505	&	-9.674	\\
-0.75	&	-6.495	&	-9.334	&	-6.495	&	-9.585	&	-6.938	&	-8.844	&	-6.495	&	-10.734	&	-6.642	&	-9.546	\\
-0.65	&	-6.574	&	-9.063	&	-6.574	&	-9.532	&	-7.012	&	-8.869	&	-6.574	&	-10.918	&	-6.713	&	-9.280	\\
-0.55	&	-6.708	&	-9.282	&	-6.709	&	-9.555	&	-7.070	&	-8.740	&	-6.708	&	-10.566	&	-6.856	&	-9.497	\\
-0.45	&	-6.795	&	-9.226	&	-6.795	&	-9.520	&	-7.151	&	-8.841	&	-6.795	&	-10.245	&	-6.943	&	-9.436	\\
-0.35	&	-6.851	&	-9.211	&	-6.851	&	-9.459	&	-7.211	&	-9.011	&	-6.851	&	-10.081	&	-6.994	&	-9.413	\\
-0.25	&	-6.900	&	-9.383	&	-6.900	&	-9.520	&	-7.268	&	-8.973	&	-6.900	&	-9.782	&	-7.047	&	-9.577	\\
-0.15	&	-7.048	&	-9.359	&	-7.048	&	-9.500	&	-7.348	&	-8.893	&	-7.048	&	-9.566	&	-7.187	&	-9.546	\\
-0.05	&	-7.162	&	-9.191	&	-7.162	&	-9.357	&	-7.463	&	-8.845	&	-7.162	&	-9.218	&	-7.305	&	-9.378	\\
0.05	&	-7.360	&	-9.191	&	-7.360	&	-9.353	&	-7.544	&	-8.846	&	-7.360	&	-8.725	&	-7.498	&	-9.379	\\
0.15	&	-7.505	&	-9.082	&	-7.504	&	-9.297	&	-7.665	&	-8.738	&	-7.505	&	-8.472	&	-7.642	&	-9.270	\\
0.25	&	-7.697	&	-8.909	&	-7.697	&	-9.088	&	-7.716	&	-8.716	&	-7.697	&	-8.293	&	-7.838	&	-9.103	\\
0.35	&	-8.115	&	-9.034	&	-8.115	&	-9.310	&	-7.878	&	-8.639	&	-8.115	&	-8.198	&	-8.224	&	-9.219	\\
0.45	&	-8.456	&	-9.016	&	-8.455	&	-9.328	&	-8.098	&	-8.628	&	-8.456	&	-8.164	&	-8.581	&	-9.203	\\
0.55	&	-8.258	&	-8.950	&	-8.258	&	-9.429	&	-8.308	&	-8.640	&	-8.258	&	-8.155	&	-8.415	&	-9.151	\\
0.65	&	-8.874	&	-9.065	&	-8.874	&	-9.497	&	-8.471	&	-8.650	&	-8.874	&	-8.180	&	-9.032	&	-9.261	\\
0.75	&	-8.659	&	-9.212	&	-8.659	&	-9.551	&	-8.692	&	-8.611	&	-8.659	&	-8.243	&	-8.839	&	-9.406	\\
0.85	&	-8.748	&	-9.440	&	-8.748	&	-9.657	&	-9.043	&	-8.567	&	-8.748	&	-8.293	&	-8.915	&	-9.629	\\
0.95	&	-9.321	&	-9.515	&	-9.321	&	-9.601	&	-8.855	&	-8.494	&	-9.321	&	-8.413	&	-9.327	&	-9.704	\\
1.05	&	-8.307	&	-9.624	&	-8.307	&	-9.735	&	-8.512	&	-8.363	&	-8.307	&	-8.577	&	-8.460	&	-9.819	\\
1.15	&	-8.021	&	-9.637	&	-8.021	&	-9.669	&	-8.164	&	-8.316	&	-8.021	&	-8.633	&	-8.169	&	-9.827	\\
1.25	&	-7.951	&	-9.609	&	-7.951	&	-9.626	&	-7.971	&	-8.212	&	-7.951	&	-8.729	&	-8.087	&	-9.799	\\
\hline
\end{tabular}
\label{tab:DTD_02}
\end{table*}

\begin{table*}
\centering
\caption{Double compact star merger r-process element yields DTD for Z=0.002. Values are logarithmic, in M$_{\odot}$ per time bin after a starburst with total mass 1 M$_{\odot}$.}
\begin{tabular}{c | c c c c c c c c c c}
\hline
log $t$	& mod. 2/4 & mod. 2/4 & mod. 5 & mod. 5 & mod. 7 & mod. 7 & mod. 12 & mod. 12 & mod. 13 & mod. 13 \\
(Gyr)	&	NSNS	&	BHNS	&	NSNS	&	BHNS	&	NSNS	&	BHNS	&	NSNS	&	BHNS	&	NSNS	&	BHNS	\\
\hline
-2.05	&	 	&	-9.634	&	 	&	-9.515	&	 	&	-8.590	&	 	&	 	&	 	&	-9.776	\\
-1.95	&	 	&	-8.359	&	 	&	-8.316	&	 	&	-8.632	&	 	&	 	&	 	&	-8.500	\\
-1.85	&	-10.899	&	-8.030	&	-10.338	&	-7.969	&	-10.847	&	-8.469	&	-10.899	&	-10.506	&	-11.194	&	-8.171	\\
-1.75	&	-11.110	&	-7.965	&	-11.291	&	-7.876	&	-10.927	&	-8.645	&	-11.110	&	-11.618	&	-11.409	&	-8.106	\\
-1.65	&	-11.275	&	-8.236	&	-11.785	&	-8.205	&	-10.627	&	-8.364	&	-11.275	&	-11.176	&	-11.542	&	-8.379	\\
-1.55	&	-11.443	&	-8.560	&	-10.473	&	-8.533	&	-9.910	&	-8.403	&	-11.443	&	-10.236	&	-11.727	&	-8.721	\\
-1.45	&	-11.779	&	-8.792	&	-10.926	&	-8.766	&	-9.334	&	-8.354	&	-11.779	&	-10.450	&	-12.055	&	-8.967	\\
-1.35	&	-9.791	&	-8.961	&	-9.746	&	-8.923	&	-9.106	&	-8.484	&	-9.791	&	-10.530	&	-9.866	&	-9.146	\\
-1.25	&	-10.419	&	-9.096	&	-10.430	&	-9.055	&	-9.162	&	-8.314	&	-10.419	&	-10.642	&	-10.674	&	-9.282	\\
-1.15	&	-9.854	&	-9.148	&	-9.865	&	-9.096	&	-8.171	&	-8.235	&	-9.854	&	-10.760	&	-9.994	&	-9.327	\\
-1.05	&	-9.304	&	-9.066	&	-9.274	&	-8.997	&	-7.968	&	-8.272	&	-9.304	&	-11.214	&	-9.419	&	-9.233	\\
-0.95	&	-9.270	&	-8.904	&	-9.248	&	-8.828	&	-8.225	&	-8.269	&	-9.270	&	-11.099	&	-9.385	&	-9.060	\\
-0.85	&	-9.190	&	-8.737	&	-9.170	&	-8.661	&	-8.514	&	-8.262	&	-9.190	&	-11.010	&	-9.317	&	-8.886	\\
-0.75	&	-9.401	&	-8.660	&	-9.286	&	-8.593	&	-8.675	&	-8.237	&	-9.401	&	-11.244	&	-9.535	&	-8.804	\\
-0.65	&	-9.556	&	-8.653	&	-9.539	&	-8.591	&	-8.841	&	-8.222	&	-9.556	&	-11.806	&	-9.688	&	-8.789	\\
-0.55	&	-9.825	&	-8.680	&	-9.900	&	-8.623	&	-8.977	&	-8.192	&	-9.825	&	-11.888	&	-9.940	&	-8.804	\\
-0.45	&	-10.113	&	-8.715	&	-10.047	&	-8.663	&	-9.184	&	-8.159	&	-10.113	&	-12.397	&	-10.217	&	-8.827	\\
-0.35	&	-10.186	&	-8.751	&	-10.867	&	-8.700	&	-9.414	&	-8.134	&	-10.186	&	-11.309	&	-10.317	&	-8.848	\\
-0.25	&	-10.625	&	-8.719	&	-11.673	&	-8.720	&	-9.624	&	-8.080	&	-10.625	&	-9.502	&	-10.757	&	-8.800	\\
-0.15	&	-11.049	&	-8.775	&	 	&	-8.729	&	-9.925	&	-8.047	&	-11.049	&	-10.340	&	-11.188	&	-8.850	\\
-0.05	&	 	&	-8.750	&	 	&	-8.737	&	-10.162	&	-7.973	&	 	&	-10.028	&	 	&	-8.811	\\
0.05	&	 	&	-8.636	&	 	&	-8.721	&	-10.421	&	-7.968	&	 	&	-9.761	&	 	&	-8.686	\\
0.15	&	 	&	-8.379	&	 	&	-8.706	&	-10.900	&	-7.859	&	 	&	-9.383	&	 	&	-8.411	\\
0.25	&	 	&	-8.683	&	 	&	-8.687	&	-11.074	&	-7.809	&	 	&	-8.773	&	 	&	-8.732	\\
0.35	&	-13.352	&	-8.388	&	 	&	-8.670	&	-11.266	&	-7.932	&	-13.352	&	-8.845	&	-13.636	&	-8.436	\\
0.45	&	-11.545	&	-8.529	&	-11.567	&	-8.658	&	-10.600	&	-7.826	&	-11.545	&	-8.827	&	-11.814	&	-8.576	\\
0.55	&	-11.276	&	-8.676	&	-10.927	&	-8.640	&	-10.242	&	-7.778	&	-11.276	&	-8.591	&	-11.541	&	-8.729	\\
0.65	&	-11.323	&	-8.870	&	-10.834	&	-8.615	&	-9.953	&	-7.930	&	-11.323	&	-8.794	&	-11.572	&	-8.912	\\
0.75	&	-11.349	&	-8.918	&	-11.026	&	-8.619	&	-9.377	&	-7.856	&	-11.349	&	-8.740	&	-11.573	&	-8.953	\\
0.85	&	-11.625	&	-8.944	&	-11.321	&	-8.639	&	-9.255	&	-7.946	&	-11.625	&	-8.863	&	-11.893	&	-8.977	\\
0.95	&	-11.436	&	-8.985	&	-11.271	&	-8.683	&	-9.217	&	-7.903	&	-11.436	&	-8.832	&	-11.671	&	-9.017	\\
1.05	&	 	&	-9.067	&	-11.675	&	-8.747	&	-9.221	&	-8.077	&	 	&	-8.822	&	 	&	-9.100	\\
1.15	&	 	&	-9.171	&	-12.231	&	-8.805	&	-9.186	&	-8.068	&	 	&	-8.685	&	 	&	-9.205	\\
1.25	&	 	&	-9.318	&	-12.867	&	-8.887	&	-9.246	&	-8.019	&	 	&	-8.758	&	 	&	-9.351	\\
\hline
\end{tabular}
\label{tab:yield_002}
\end{table*}

\begin{table*}
\centering
\caption{Double compact star merger r-process element yields DTD for Z=0.002. Values are logarithmic, in M$_{\odot}$ per time bin after a starburst with total mass 1 M$_{\odot}$.}
\begin{tabular}{c | c c c c c c c c c c}
\hline
log $t$	& mod. 2/4 & mod. 2/4 & mod. 5 & mod. 5 & mod. 7 & mod. 7 & mod. 12 & mod. 12 & mod. 13 & mod. 13 \\
(Gyr)	&	NSNS	&	BHNS	&	NSNS	&	BHNS	&	NSNS	&	BHNS	&	NSNS	&	BHNS	&	NSNS	&	BHNS	\\
\hline
-1.85	&	 	&	-11.235	&	 	&	-11.235	&	 	&	-9.275	&	 	&	 	&	 	&	-11.381	\\
-1.75	&	 	&	-10.316	&	 	&	-10.317	&	 	&	-9.314	&	 	&	 	&	 	&	-10.456	\\
-1.65	&	 	&	-10.164	&	 	&	-10.204	&	 	&	-9.339	&	 	&	-11.375	&	 	&	-10.308	\\
-1.55	&	-11.156	&	-10.082	&	-11.153	&	-10.116	&	-9.710	&	-9.671	&	-11.156	&	-11.216	&	-11.464	&	-10.219	\\
-1.45	&	-11.556	&	-10.326	&	-11.587	&	-10.387	&	-9.801	&	-9.857	&	-11.556	&	-11.212	&	-11.864	&	-10.497	\\
-1.35	&	-11.173	&	-10.687	&	-11.173	&	-10.796	&	-9.946	&	-9.986	&	-11.173	&	-11.909	&	-11.464	&	-10.856	\\
-1.25	&	-10.649	&	-10.726	&	-10.647	&	-10.895	&	-9.868	&	-10.218	&	-10.649	&	-12.600	&	-10.922	&	-10.913	\\
-1.15	&	-8.635	&	-10.777	&	-8.635	&	-10.930	&	-8.556	&	-10.243	&	-8.635	&	-12.161	&	-8.765	&	-10.977	\\
-1.05	&	-8.125	&	-10.945	&	-8.125	&	-10.987	&	-8.274	&	-10.196	&	-8.125	&	-12.053	&	-8.251	&	-11.155	\\
-0.95	&	-8.158	&	-10.903	&	-8.158	&	-11.129	&	-8.529	&	-10.246	&	-8.158	&	-12.230	&	-8.300	&	-11.113	\\
-0.85	&	-8.322	&	-10.895	&	-8.322	&	-11.272	&	-8.797	&	-10.388	&	-8.322	&	-12.551	&	-8.459	&	-11.115	\\
-0.75	&	-8.449	&	-10.773	&	-8.449	&	-11.135	&	-8.885	&	-10.359	&	-8.449	&	-12.409	&	-8.595	&	-10.991	\\
-0.65	&	-8.526	&	-10.452	&	-8.526	&	-11.075	&	-8.959	&	-10.391	&	-8.526	&	-12.593	&	-8.664	&	-10.674	\\
-0.55	&	-8.659	&	-10.718	&	-8.659	&	-11.108	&	-9.014	&	-10.189	&	-8.659	&	-12.167	&	-8.807	&	-10.938	\\
-0.45	&	-8.742	&	-10.664	&	-8.742	&	-11.110	&	-9.095	&	-10.255	&	-8.742	&	-11.819	&	-8.890	&	-10.881	\\
-0.35	&	-8.799	&	-10.668	&	-8.799	&	-11.065	&	-9.154	&	-10.434	&	-8.799	&	-11.633	&	-8.942	&	-10.879	\\
-0.25	&	-8.847	&	-10.859	&	-8.847	&	-11.074	&	-9.212	&	-10.421	&	-8.847	&	-11.309	&	-8.994	&	-11.059	\\
-0.15	&	-8.993	&	-10.835	&	-8.992	&	-11.050	&	-9.292	&	-10.311	&	-8.993	&	-11.103	&	-9.131	&	-11.028	\\
-0.05	&	-9.109	&	-10.662	&	-9.109	&	-10.913	&	-9.406	&	-10.240	&	-9.109	&	-10.728	&	-9.252	&	-10.857	\\
0.05	&	-9.307	&	-10.651	&	-9.307	&	-10.888	&	-9.488	&	-10.242	&	-9.307	&	-10.222	&	-9.445	&	-10.842	\\
0.15	&	-9.453	&	-10.529	&	-9.453	&	-10.843	&	-9.613	&	-10.130	&	-9.453	&	-9.959	&	-9.591	&	-10.721	\\
0.25	&	-9.645	&	-10.375	&	-9.645	&	-10.644	&	-9.667	&	-10.108	&	-9.645	&	-9.780	&	-9.785	&	-10.571	\\
0.35	&	-10.067	&	-10.449	&	-10.067	&	-10.825	&	-9.828	&	-10.033	&	-10.067	&	-9.689	&	-10.177	&	-10.636	\\
0.45	&	-10.410	&	-10.434	&	-10.409	&	-10.869	&	-10.050	&	-10.008	&	-10.410	&	-9.663	&	-10.536	&	-10.623	\\
0.55	&	-10.214	&	-10.325	&	-10.214	&	-10.935	&	-10.262	&	-10.001	&	-10.214	&	-9.662	&	-10.370	&	-10.529	\\
0.65	&	-10.828	&	-10.441	&	-10.828	&	-10.984	&	-10.425	&	-10.006	&	-10.828	&	-9.698	&	-10.986	&	-10.640	\\
0.75	&	-10.614	&	-10.608	&	-10.614	&	-11.057	&	-10.646	&	-9.956	&	-10.614	&	-9.769	&	-10.794	&	-10.806	\\
0.85	&	-10.703	&	-10.865	&	-10.703	&	-11.164	&	-10.997	&	-9.896	&	-10.703	&	-9.826	&	-10.870	&	-11.056	\\
0.95	&	-11.254	&	-10.969	&	-11.254	&	-11.090	&	-10.795	&	-9.807	&	-11.254	&	-9.952	&	-11.260	&	-11.159	\\
1.05	&	-10.244	&	-11.080	&	-10.244	&	-11.239	&	-10.442	&	-9.674	&	-10.244	&	-10.126	&	-10.396	&	-11.274	\\
1.15	&	-9.953	&	-11.125	&	-9.953	&	-11.174	&	-10.101	&	-9.607	&	-9.953	&	-10.180	&	-10.100	&	-11.314	\\
1.25	&	-9.887	&	-11.113	&	-9.887	&	-11.136	&	-9.911	&	-9.497	&	-9.887	&	-10.281	&	-10.021	&	-11.301	\\
\hline
\end{tabular}
\label{tab:yield_02}
\end{table*}

\begin{table*}
\centering
\caption{The different models for which results are given in the present paper (rejected ones in italic). See text for the definition of parameters and their values, and see Mennekens \& Vanbeveren (2014) for details about these and other rejected models.}
\begin{tabular}{c c c c c c c c c c c c}
\hline
Model &	ECSN & $\beta$ & $\alpha_{CE}$ & Fallback & Period & LBV & BH & avg. NS \\
 & & & & model & distribution & & kicks & kick (km/s) \\
\hline
2/4 & On & 1 & 0.5 & Rapid & Flat & On & On & 450 \\
5 & On & 1 & 0.5 & Delayed & Flat & On & On & 450 \\
\emph{6} & \emph{On} & \emph{1} & \emph{1} & \emph{Rapid} & \emph{Flat} & \emph{On} & \emph{On} & \emph{450} \\
7 & Off & 1 & 1 & Rapid & Flat & On & On & 450 \\
12 & On & 1 & 0.5 & Rapid & Flat & Off & On & 450 \\
13 & On & 1 & 0.5 & Rapid & Alternative & On & On & 450 \\
\emph{19} & \emph{On} & \emph{1} & \emph{5} & \emph{Rapid} & \emph{Flat} & \emph{On} & \emph{On} & \emph{450} \\
\emph{22} & \emph{On} & \emph{0.5} & \emph{1} & \emph{Rapid} & \emph{Flat} & \emph{On} & \emph{On} & \emph{450} \\
\hline
\end{tabular}
\label{tab:models}
\end{table*}

Tables 1 and 2 give the DTD of NS+NS and of NS+BH mergers for the non-rejected models as discussed above. Table 3 reminds the reader of the assumptions behind these models, which are laid out in more detail in Paper I. The calculated rates are for a starburst with total mass = 1 M$_{\odot}$, and a binary frequency = 100\%. It is then obvious to estimate the rates for any starburst whatever the binary frequency. In Tables 3 and 4 (also available online at the same address) we give the corresponding DTDs of the r-process yields by linking the calculations of Korobkin et al. (2012) with our predicted temporal evolution of the population of compact star mergers. Figure 3 shows the DTD for model 2. All the models retained in the present paper give very similar DTDs.  The DTDs illustrate that NS+NS or/and NS+BH mergers show up typically 10-30 Myr after the onset of the starburst whereas the merger rates depend critically on the metallicity Z. Figure 5 is similar to Figure 3 but for a binary period distribution proposed by Sana et al. (2012). As can be noticed the difference is moderate.

\begin{figure}[]
\centering
   \includegraphics[width=8.4cm]{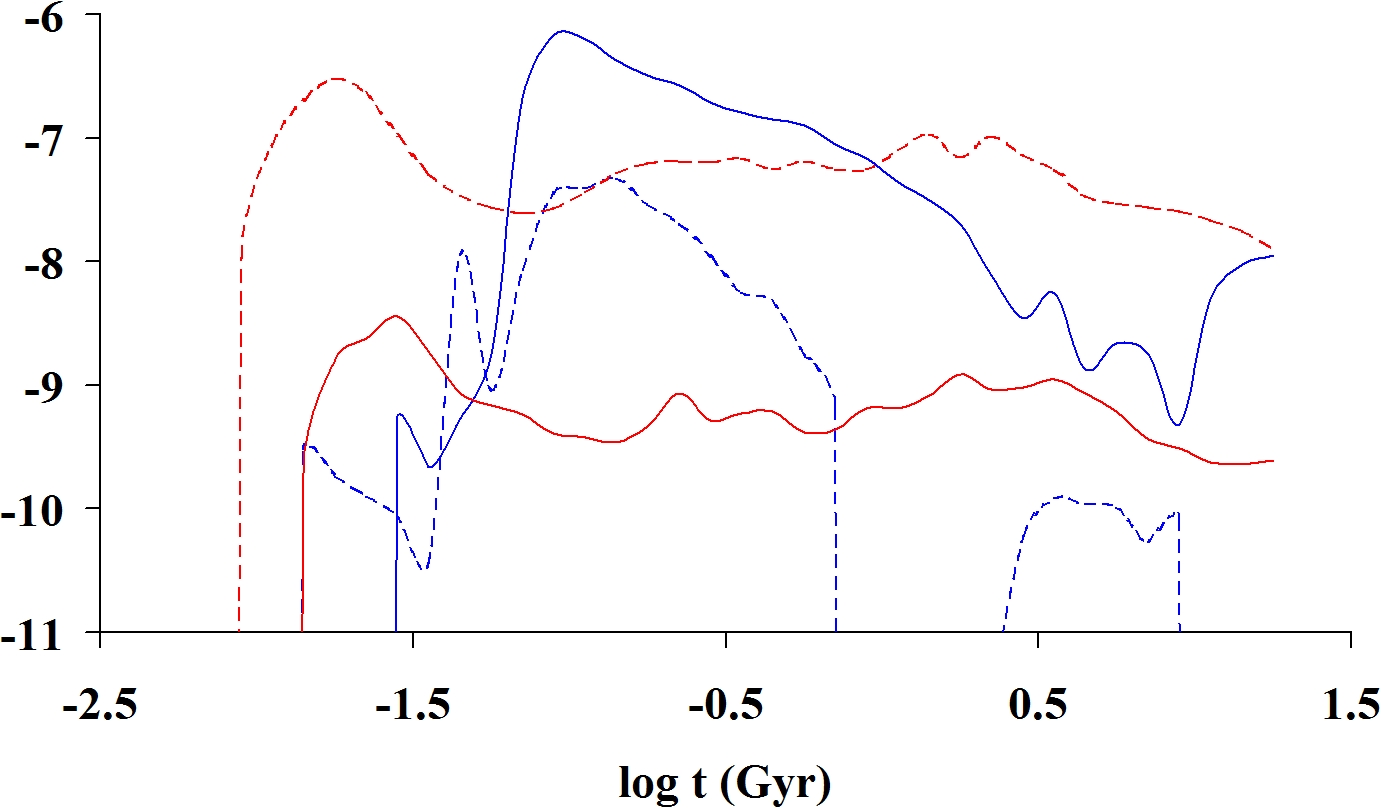}
     \caption{DTD of double compact star mergers obtained using model 2/4, in logarithmic number of events per time bin after a starburst with total mass 1 M$_{\odot}$. Blue lines indicate NSNS mergers, red lines BHNS mergers. Dashed lines indicate Z=0.002, solid lines Z=0.02.}
     \label{fig:3}
\end{figure}

\begin{figure}[]
\centering
   \includegraphics[width=8.4cm]{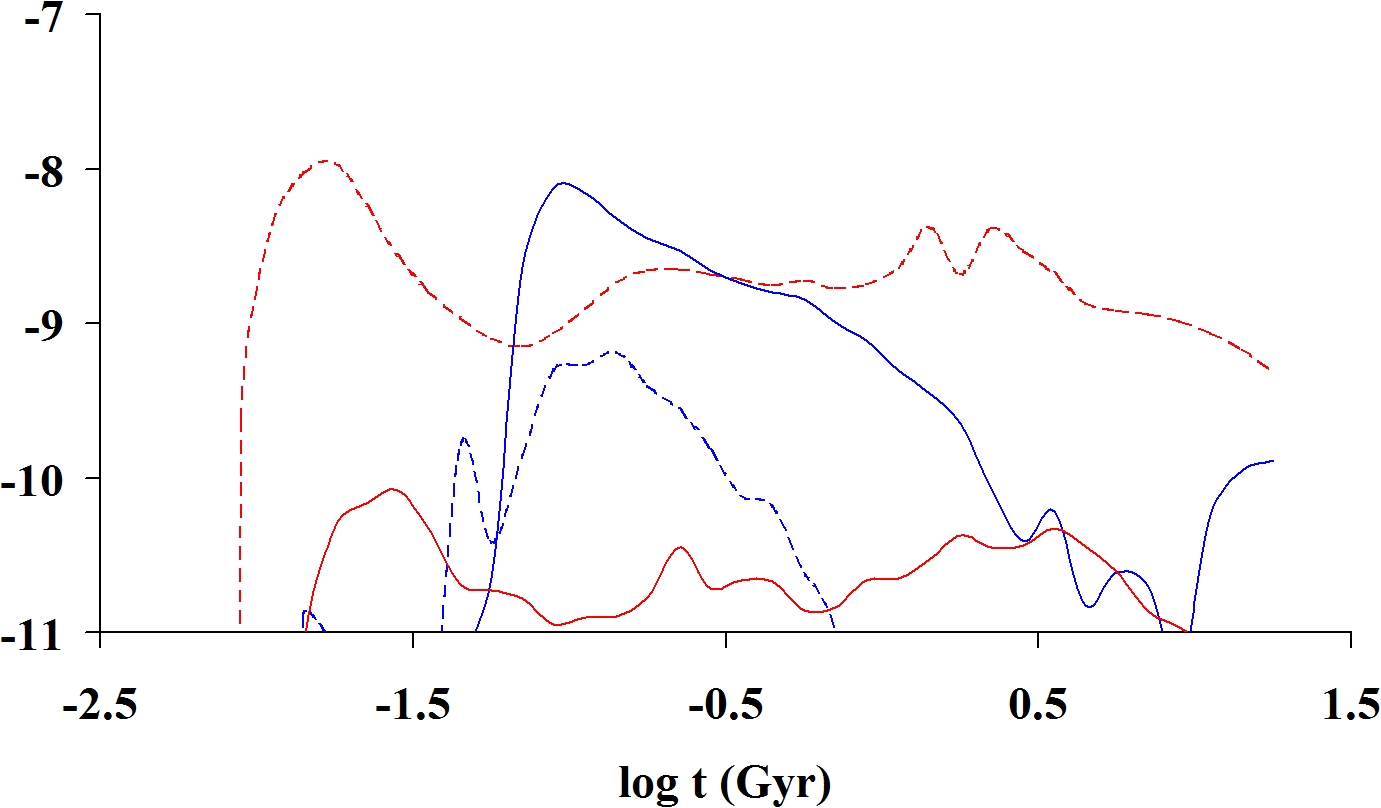}
     \caption{DTD of double compact star merger r-process element yields obtained using model 2/4, in logarithmic M$_{\odot}$ per time bin after a starburst with total mass 1 M$_{\odot}$. Blue lines indicate NSNS mergers, red lines BHNS mergers. Dashed lines indicate Z=0.002, solid lines Z=0.02.}
     \label{fig:4}
\end{figure}

\begin{figure}[]
\centering
   \includegraphics[width=8.4cm]{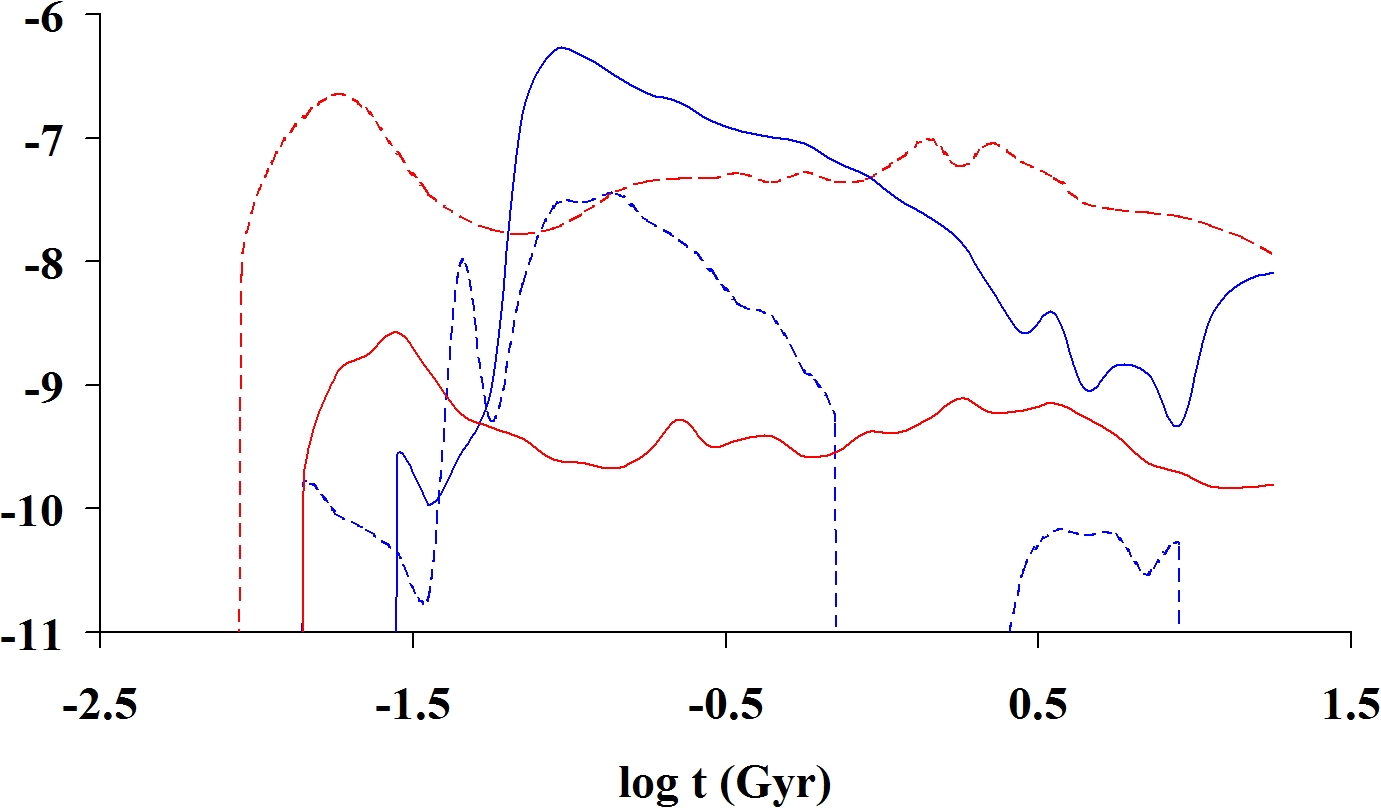}
     \caption{DTD of double compact star mergers obtained using model 13, in logarithmic number of events per time bin after a starburst with total mass 1 M$_{\odot}$. Blue lines indicate NSNS mergers, red lines BHNS mergers. Dashed lines indicate Z=0.002, solid lines Z=0.02.}
     \label{fig:5}
\end{figure}

\section{Comparison to earlier work}

Many research groups studied the populations of massive compact star pairs and some of them discuss the delay time distributions. One of the earliest papers is the one of Tutukov and Yungelson (1994), a paper published in the era where it was believed that the supernova explosion of a massive star is spherically symmetric. The results of Tutukov and Yungelson are calculated with the latter assumption. For more recent work (with in particular a non-spherical treatment of the SN explosion) we like to refer to De Donder and Vanbeveren (2004) and Dominik et al. (2012) and references therein. In all these studies the delay time is defined as the orbital decay time between the formation of the double compact star binary and the moment of merging and obviously the published distributions describe the distributions of these decay times. However, this is not the one needed in order to study the effect of massive binaries on the r-process element enrichment of galaxies with a code that does not explicitly account for binaries. What is really needed is the delay time distribution starting from a binary population on the zero age main sequence up to the double compact star merger, i.e. the delay time must include the evolutionary timescale of the progenitor binary. The DTDs presented here account for this and therefore they can hardly be compared to the DTDs presented in earlier work.

\section{Conclusion}

The theoretically predicted delayed time distribution of NS+NS and of NS+BH mergers depends on binary parameters whose values are poorly constrained from first principles. However, a careful comparison with observations allows to promote a most plausible DTD that can easily be implemented in a galactic chemical code in this way avoiding the need to add a full binary population model. This may in particular be very useful in order to study the r-process element enrichment of dwarf galaxies, building blocks of spirals. Our results reveal that the galactic chemo-binary models that predict a r-process element enrichment that is not too large compared to observations, all have a galactic double neutron star merger rate at present $\leq$ 10/Myr (corresponding to a combined NS+NS and NS+BH merger rate $\leq$ 20/Myr). The latter has obviously very important consequences for the analysis and interpretation of the results of aLIGO that will become available in the very near future.

\begin{acknowledgements}
      We thank the anonymous referee for valuable comments.
\end{acknowledgements}

\end{document}